\documentclass{article}
\usepackage {graphicx}
\begin{document}
\baselineskip .75cm 
\begin{titlepage}
\title{\bf Fully nonlinear excitations of non-Abelian plasma}       
\author{Vishnu M. Bannur  \\
{\it Department of Physics}, \\  
{\it University of Calicut, Kerala-673 635, India.} }   
\maketitle
\begin{abstract}

We investigate fully nonlinear, non-Abelian excitations of quark-antiquark plasma, using relativistic fluid theory in cold plasma approximation. There are mainly three important nonlinearities, coming from various sources such as non-Abelian interactions of Yang-Mills (YM) fields, Wong's color dynamics and plasma nonlinearity, in our model. By neglecting nonlinearities due to plasma and color dynamics we get back the earlier results of Blaizot {\it et. al.} \cite{bl.1}. Similarly, by neglecting YM fields nonlinearity and plasma nonlinearity, it reduces to the model of Gupta {\it et. al.} \cite{su.1}. Thus we have the most general non-Abelian mode of quark-gluon plasma (QGP). Further, our model resembles the problem of propagation of laser beam through relativistic plasma \cite{ka.1} in the absence of all non-Abelian interactions.   
 
\end{abstract}
\vspace{1cm}
                                                         
\noindent
{\bf PACS Nos :} 12.38.Mh, 05.45.-a, 52.35.Mw, 52.27.Ny \\
{\bf Keywords :} Quark-gluon plasma, Non-Abelian, Nonlinearity.  
\end{titlepage}
\section{Introduction :}

QGP is a quasi-color-neutral gas of quarks and gluons which exhibits collective behaviour. It is expected to be formed in relativistic heavy ion collisions (RHICs) experiments, deep inside the neutron star and might have formed in early universe. A study of collective excitations of QGP are important to diagnose various parameters and signatures of QGP. It is also proposed that the chaotic collective modes of QGP gives an estimate of thermalization of QGP in RHICs \cite{mu.1}. From the extensive study of electrodynamics plasma, we know that there exists various linear and nonlinear excitations in it, governed by electrodynamic interactions which is an Abelian gauge theory. Here, in QGP, also we expect similar linear and nonlinear modes, but modified by the non-Abelian interactions, which itself is nonlinear. Therefore, in QGP, there is two types of nonlinear effects, one coming from usual plasma nonlinearity and other, from non-Abelian effects. Nonlinear solutions of non-Abelian or Yang-Mills (YM) theory is studied extensively by Matinyan {\it et. al.} without plasma, but with Higgs order phase \cite{ma.1}. Later, these studies were extended to QGP by Blaizot {\it et. al.}, but without plasma nonlinearity and Wong's color dynamics \cite{wa.1} which is another non-Abelian effect in QGP. 
Various periodic, quasi-periodic, chaotic nonlinear modes and transition from order to chaos by plasma collective effects were studied. A study of stabilization of QCD vacuum instability by plasma collective modes were studied earlier in \cite{ba.1}.   
There is another group of work in these lines by Gupta {\it et. al.}, but without the nonlinearity of YM fields  and plasma nonlinearity. Nonlinear or Non-Abelian effects come from the Wong's color dynamics. Here, we present a fully nonlinear, non-Abelian excitations, including all nonlinearities: plasma nonlinearity, YM fields nonlinearity and color dynamics nonlinearity.       

\section{Relativistic fluid theory of QGP:} 

The relativistic fluid set of equations, in cold plasma limit, is given by \cite{se.1} 
\begin{equation}
m \frac{d u^{\mu}}{d {\tau}} = g I_a G^{\mu \nu}_a u_{\nu} \,\,, 
\end{equation}
the equation of motion, where $m$ the mass, $\tau$ the proper time, $g$ the coupling constant, $a$ the color index, $u^{\nu}$ 4-velocity, which is also the fluid velocity in cold plasma limit, and $G^{\mu \nu}_a$ the field tensor, defined as, 
\begin{equation}
G^{\mu \nu}_a = \partial^{\mu} A^{\nu}_a - \partial^{\nu} A^{\mu}_a 
+ g \epsilon_{abc} A^{\mu}_b A^{\nu}_c \,\,,    
\end{equation}
in terms of 4-vector potentials $A^{\mu}_a$. 
$I_a$ is the dynamical color charges which obey Wong's equation,   
\begin{equation}
\frac{d I_a}{d \tau} = - g \epsilon_{abc} u^{\mu} A_{\mu b} I_c \,\,.   
\end{equation}
The vector potentials are obtained from the Yang-Mills field equation, 
\begin{equation}
\partial_{\mu} G^{\mu \nu}_a + g \epsilon_{abc} A_{\mu b} G^{\mu \nu}_c = J^{\nu}_a \,\,,  
\end{equation}
where $J^{\nu}_a$ is the 4-vector color current produced by various species in plasma with color charges, such as quarks, antiquarks and gluons. For simplicity, here in our analysis, we consider quark-antiquark plasma and the current density is given by 
\begin{equation}
J^{\nu}_a = \sum_{species} n g I_a u^{\nu} \,\,, 
\end{equation}
where $n$ is the density of each species, determined by the continuity equation, 
\begin{equation}
\partial _{\mu} (n u^{\mu}) = 0 \,\,.  
\end{equation} 
In general, these are very complicated, coupled, nonlinear equations to be solved and hence one goes for approximations, such as moving frame ansatz \cite{ka.1}, space-homogeneous solutions, so on, to look for special solutions. Following Blaizot {\it et. al.} \cite{bl.1}, let us consider the homogeneous solutions of our set of equations, few of them may be easily solved. The continuity equation for each species may be integrated and we get 
\begin{equation}
n(t) u^0 (t) = constant = n_0 u^0_0 \,\,, 
\end{equation}
where $n_0$ and $u^0_0$ are the density and 0-component of fluid velocity at equilibrium. We also chose a gauge $A^0_a = 0$ and the spatial part of the equation of motion may be easily integrated to get 
\begin{equation}
u^j = - \frac{g I_a A^j_a}{m} \,\, , \label{eq:ui} 
\end{equation}
with the assumption that, at equilibrium, the plasma is at rest. The 0-component fluid velocity is given by 
\begin{equation}
u^0 = \sqrt{1 + u_j^2} \,\,,
\end{equation}   
and hence $u^0_0 = 1$. 
Similarly, the spatial part of the field equations gives 
\begin{equation}
\ddot{A^i_a} + g^2 \left[ (A^j_b A^j_b) A^i_a -  (A^j_b A^j_a) A^i_b \right] = 
g \sum n_0 I_a \frac{u^i}{u^0} \,\, ,
\end{equation}
and temperal component gives,
\begin{equation}
\sum n_0 I_a = \epsilon_{abc} A^i_b \dot{A^i_c} \,\,, \label{eq:isum}  
\end{equation} 
where dot means differentiation with respect to time. Finally, the color dynamics equation reduces to 
\begin{equation}
\dot{I_a} = g \epsilon_{abc} \frac{u^i}{u^0} A^i_b I_c \,\, . 
\end{equation}
For further simplification, let us use hedgehog ansatz where the color directions are taken to be along the spatial direction and redefine variables as 
\begin{equation}
X \equiv \frac{g I_0 A_{x1}}{m};\;\; Y \equiv \frac{g I_0 A_{y2}}{m};\;\; Z \equiv \frac{g I_0 A_{z3}}{m};\;\; , 
\end{equation}
and rescaling time and color charges as   
\begin{equation}
 t \rightarrow \frac{m}{I_0} t \;\;\mbox{and}\;\; I_a \rightarrow \frac{I_a}{I_0} \,\,, 
\end{equation}
where $I_0$ is introduced to normalize $I_a I_a = 1$, which is one of the constant of motion as can be seen from the equation for color dynamics. Further, from Eq. (\ref{eq:isum}), $I_a$ of second species (antiquarks) is opposite to that of first species (quarks) and hence $I_{2a} = - I_{1a} \equiv - I_a$. In terms of redefined variables, our simplified set of equations becomes, 
\begin{equation}
\ddot{X} + (Y^2 + Z^2) X = - \epsilon \,I_{x}^2 \frac{X}{\sqrt{1 + (I_{x} X)^2 + (I_{y} Y)^2 +  (I_{z} Z)^2}} \,\,, \label{eq:ymh}   
\end{equation}
and 
\begin{equation}
\dot{I}_{x} = - \frac{I_{y} I_{z} (Y^2 - Z^2)}{ \sqrt{1 + (I_{x} X)^2 + (I_{y} Y)^2 +  (I_{z} Z)^2}}\,\,,  
\end{equation}
and similar equations for $y$ and $z$ components which may be obtained by cyclic change among $x$, $y$ and $z$. The parameter $\epsilon \equiv \frac{2 \omega_p^2 I_0^2}{m^2}$, where the plasma frequency $\omega_p^2 \equiv \frac{n_0 g^2 I_0^2}{m}$. Above set of equations has two immediate constant of motions, namely, $I_{a} I_{a} = 1$ and 
\begin{equation}
(\dot{X}^2 + \dot{Y}^2 + \dot{Z}^2)/2 + (X^2 Y^2 + Y^2 Z^2 + Z^2 X^2)/2 + 
\epsilon \, \sqrt{1 + (I_{x} X)^2 + (I_{y} Y)^2 +  (I_{z} Z)^2} = E\,\,,    
\end{equation}     
the energy. These approximate set of equations retains all important aspects of QGP such as YM nonlinearity, plasma nonlinearity and color dynamics nonlinearity. In the earlier calculations of Blaizot {\it et. al.} \cite{bl.1}, the color dynamics is neglected and in \cite{su.1}, YM nonlinearity is dropped out. 

In order to extract the results of Blaizot {\it et. al.}, let us assume that color charge $I_a$ are constant and then, Eq. (\ref{eq:ymh}) reduces to,
\begin{equation}
\ddot{X} + (Y^2 + Z^2) X = - \epsilon \frac{1}{3} 
\frac{X}{\sqrt{1 + (X^2 + Y^2 + Z^2)/3}} \,\,,  \label{eq:ymha}   
\end{equation}     
where the square root term is the plasma nonlinearity, coming from the relativistic treatment just like in \cite{ka.1}.  
Further, expanding the plasma nonlinearity term up to $3^{rd}$ order in vector potential gives,
\begin{equation}
\ddot{X} + (1 - \frac{\epsilon}{18}) (Y^2 + Z^2) X - \frac{\epsilon}{18} X^3  
+\frac{\epsilon}{3} X = 0 \,\,,   
\end{equation}     
which is similar to that of Blaizot {\it et. al.}, except with few new terms containing a   $ (- \frac{\epsilon}{18})$. This new terms may lead to additional new features like chaotic scattering \cite{ko.1}. This model without these new additional term was studied extensively by Matinyan {\it et. al.} \cite{ma.1} and Blaizot {\it et. al.} \cite{bl.1}.  

Let us look for some other new solutions of our model Eq. (\ref{eq:ymha}). For example, a special solution with $Z=0$ leads to, 
\begin{equation}
\ddot{X} + Y^2 X = - \epsilon \,\frac{1}{3}\, 
\frac{X}{\sqrt{1 + (X^2 + Y^2)/3}} \,\,,     
\end{equation}     
for $X$ and a similar equations for $Y$ with $X$ and $Y$ interchanged. It differs from similar work by Matinyan {\it et. al.}, Blaizot {\it et. al.} because we have kept the plasma nonlinearity also. Our numerical study shows that the plasma nonlinearity enhances the chaos and therefore, increases the order to chaos transition parameter defined in \cite{ma.1}, which will be discussed later. 

Next let us look at another special solution with $X = Y = Z$ of our general equation Eq. (\ref{eq:ymh}) and we get   
\begin{equation}
\ddot{X} + 2\,X^3 = - \epsilon \,\frac{1}{3} \,
\frac{X}{\sqrt{1 + X^2}} \,\,, \label{eq:ymhb}    
\end{equation}     
which describes a nonlinear oscillation. It is the more general nonlinear oscillation, including the plasma nonlinearity, than the elliptic functions $Cn$ discussed in \cite{ma.1,bl.1}. It is easy to see that on neglecting the plasma nonlinearity, we get back $Cn$ or $Sn$, depending on the strength of non-Abelian parameter compared to plasma frequency. It is interesting to note that above mode is an exact solution of QGP because for $X = Y = Z$, the color dynamics equation shows that color charges are constant. 
           

\section{Results:} 

The most general set of equations of our model are nonlinear, coupled equations and may not be easy to solve. So we have made an approximation, known as hedgehog ansatz and reduced the number of equations to be solved, but having all non-Abelian and nonlinear features. From these simplified set of equations, we may get the results of all other earlier works in this field. For example, in Fig. 1, we plotted the Poincare section of our model with the approximation that the dynamical color charges are constant and $Z=0$, Eq. (\ref{eq:ymha}). Figures 1(a) and 1(b) are for the system without plasma nonlinearity and found that the regular orbits seen in Fig. 1(a) 
for $\epsilon = 5$ disappears at the critical value of $\epsilon = 2$, 
Fig. 1(b) and hence chaotic. Similar figures with plasma nonlinearity shows changes from ordered orbit islands for $\epsilon = 8.15$, Fig. 1(c), into chaotic motion for $\epsilon = 6$. Therefore, the critical value of $\epsilon$ for the order-to-chaos transition is higher with plasma nonlinearity. The chaos seen with $\epsilon = 6.0$ with plasma nonlinearity develops islands of ordered motion without plasma nonlinearity and we need smaller $\epsilon$ ($\epsilon = 2$) to have chaos. Hence the plasma nonlinearity enhances the chaos which is an additional new feature compared to the results of Blaizot {\it et. al.} \cite{bl.1}. Another special solution of our model with $I_a = $ constant is $X = Y = Z$, Eq. (\ref{eq:ymhb}), which is not an elliptic functions as in \cite{bl.1}, but little more general nonlinear oscillation. 


Next, in Fig. 2, we plotted the general solutions of our model, as an example $u_x$ (Eq. (\ref{eq:ui})),  with hedgehog ansatz  for different values of $\epsilon$ with the same initial conditions and we see that as the $\epsilon$ decreases the system becomes more and more chaotic, which is, qualitatively, similar to the results of Gupta {\it et. al.}. For a large $\epsilon$, say,  $\epsilon = 100$ (Fig.1 (a)), the amplitude of oscillations is small and the YM nonlinearity and plasma nonlinearity may be negligible and hence it is just the Abelian oscillations, modulated by color dynamics. As $\epsilon$ decreases, amplitude increases and all nonlinearities due to YM fields, color dynamics and plasma nonlinearity come into play and drive the system to chaotic motion as can be seen from Figs. 2(b) to 2(c) with intermittent oscillations. For $\epsilon = 0$, Fig. 2(d), the chaotic oscillations is mainly due to YM nonlinearity. Similar behaviour is also seen in the other components of velocity.         
            
\section{Conclusions:} 

We have studied fully nonlinear, non-Abelian excitations of quark-antiquark plasma using relativistic fluid theory. It exhibits new features like a special nonlinear oscillation, different from elliptic functions, and enhancement of chaos. Further, we have found that on neglecting color dynamics and plasma nonlinearity, we get back the results of Blaizot {\it et. al.} \cite{bl.1} and by neglecting YM fields nonlinearity and plasma nonlinearity, we obtain the results of Gupta {\it et. al.} \cite{su.1}. Hence we have the most general nonlinear, non-Abelian modes of QGP. In general, all three nonlinearities are always there in the system. For small amplitude excitations ($|X|$, $|Y|$ and $|Z|$ are $\ll 1$), YM field nonlinearity and plasma nonlinearity may be negligible like in \cite{su.1}. At present, we don't know the appropriate limit to get the results of \cite{bl.1}, since the nonlinearity due color dynamics is always there irrespective of the strength of the excitations and the plasma nonlinearity also will be there due to relativistic treatment.   
On neglecting all non-Abelian nonlinearities our model resembles the model of laser propagation through relativistic plasma \cite{ka.1}.

\newpage 
{\bf Figure Caption:} \\[1cm] 

\begin{figure}
\caption {Poincare sections of our model (with $Z=0$ and $I_a = $ constant) without plasma nonlinearity (Figs. (a) for $\epsilon = 5$ and (b) for $\epsilon = 2$), and with plasma nonlinearity (Figs. (c) for $\epsilon = 8.15$ and (d) for $\epsilon=6$).}
\label{fig 1}
\vspace{.75cm}

\caption {Exact numerical solutions of our model, as an example $u_x$, for different values of $\epsilon$ with the same initial conditions (Figs. (a) for $\epsilon = 100$, (b) for $\epsilon = 20$, (c) for $\epsilon = 2$ and (d) for $\epsilon = 0$).} 
\label{fig 2}
\vspace{.75cm}
\end{figure}
\end{document}